# DEEP IMPACT MISSION TO TEMPEL 1

# FAVOURS NEW EXPLOSIVE COSMOGONY OF COMETS

*E.M.Drobyshevski, E.A.Kumzerova, and A.A.Schmidt*

A.F.Ioffe Physico-Technical Institute, Russian Academy of Science,
194021 St.Petersburg, Russia
E-mail: emdrob@mail.ioffe.ru

**ABSTRACT**
The assumption that short-period (SP) comets are fragments of massive icy envelopes of Ganymede-like bodies saturated by products of ice electrolysis that underwent global explosions provides a plausible explanation of all known manifestations of comets, including the jet character of outflows, the presence of ions in the vicinity of the nucleus, the bursts and splitting of cometary nuclei, etc., with solar radiation initiating burning of the products of electrolysis in the nucleus.
As shown persuasively by numerical simulation carried out in hydrodynamic approximation, the shock wave initiated by the Deep Impact (DI) impactor in the cometary ice saturated originally by the electrolysis products $2H_2 + O_2$ is capable of activating under certain conditions exothermal reactions (of the type $O_2 + H_2$ + organics $\rightarrow H_2O + CO + HCN$ + other products of incomplete burning of organics including its light and heavy pyrolyzed compounds, soot, etc.), which will slow down shock wave damping (forced detonation) and increase many times the energy release. As a result, the measured energetics of ejections and outflows from the crater have to exceed the DI energetics. Analysis of different clusters of the DI experiment data confirms these conclusions and expectations and thus it favours the planetary origin of comets.

## 1. INTRODUCTION. DEEP IMPACT ACTIVE EXPERIMENT

The DI mission culminated on 4 July, 2005, when the self-guided impactor of 370 kg in weight with a characteristic size of ~1m, a half of the mass was copper, struck the comet Tempel 1 nucleus at velocity $u_{imp}$ = 10.3 km/s. The impact energy was $E_0$ =19.6 GJ. The collision process and its consequences were observed by a flyby "mother" spacecraft 500 km distant, as well as by numerous space and ground-based telescopes (A'Hearn et al. 2005b; Meech et al. 2005).

  The DI is unprecedented in the history of space missions because this experiment is active. Its basic and natural motivation consisted in our practically total ignorance of the parameters of comet nuclei, such as their mass, density, composition, surface and internal structure, characteristics of the matter strength and so on (A'Hearn et al. 2005a; Belton et al. 2005). Really, the majority of existent estimates of these parameters critically depends on the models accepted, that in their turn, are based on the old speculative concepts on the comet nuclei as a by-product of the planet formation in the pre-planetary gas-dust cloud 4.5 Gy ago (A'Hearn et al. 2005a). In spite of these notions are frequently decorated by rich clothes of mathematical simulation, they turn out to be helpless in explanation of many old and, moreover, newly discovered facts and phenomena, let alone prediction of them. In this field, there is a wonderful wealth of *ad hoc* hypotheses invented for explaining separate facts, and very often an inconvenient factual material is simply ignored. Even in getting a cursory view, some wrenches and contradictions in many facts' interpretation from the positions of the traditional paradigms are conspicuous. In such a situation, even the natural intention to satisfy the conservation laws plays sometimes an insidious role of the Procrustean bed.

  In the present paper, we shall try to show that the New Explosive (Eruptive) Cosmogony (NEC) of comets and other minor bodies in the Solar system offers a non-contradictory explanation of the DI results. NEC is based on a self-consistent assumption that the vast majority of SP comets are actually products of global explosions of icy envelopes of Ganymede-like bodies saturated with products of the volumetric electrolysis of ice. Sec. 2 will expose briefly the main statements and achievements of the NEC. Sec. 3 will provide a short outline of present views concerning impact crater formation. Other conditions being equal, the crater volume is practically proportional to the energy released at the impact. If, as predicted by NEC, cometary ices do indeed contain, besides primitive organics and rock inclusions, products of the electrolysis of ice, more specifically, $O_2$ and $H_2$, dissolved in the form of clathrates, impact may initiate at the very least a non-self-sustained (forced) detonation (and thereafter, combustion), i.e., an additional release of energy. Calculations of the DI-excited shock wave with an additional energy release made in hydrodynamic approximation (Sec. 4) suggest that such an energy addition may, under reasonable assumptions, exceed noticeably $E_0$, which would increase correspondingly the crater size. Sec. 5 suggests also some other consequences of the activation of an internal chemical source of energy. Among them are specific features in chemical composition and long duration and excessive energetics of ejections and outflows of the dusty gas from the crater, as well as even a possibility for the nucleus to break up into fragments with kinetic energy $\geq E_0$. (Contents of Secs. 2-5 were published by us earlier – one and a half month before DI (see astro-ph/0505377 in www.arXiv.org)). In Sec. 6 the main physical results of DI are listed and analyzed. It is shown that an interpretation proposed by a number of authors sometimes contains contradictions that vanish if one leans on the NEC inferences. In the Conclusion (Sec. 7), we note again that all the accessible DI results, when considered from the NEC position, combine into the unified non-contradictory and harmonic picture. In this picture, there hardly is a place for the conventional condensation-sublimation approach on the origin of comets and nature of their activity. Considerable part of DI data rather contradicts it. An alternative to the former paradigm is the planetary origin of comets.

## 2. ON THE ORIGIN OF COMETS

Discussions bearing on the goals and possible results of the DI mission are replete with statements that they should permit us to take a look into the past of the Solar System and formulate a judgment of the primordial material of which all its bodies had been made, including the comets themselves (e.g., Belton et al. 2005, and refs. therein). The latter are considered to be nothing more than building rubble, namely, agglomerates of rocks and ices of volatile compounds left over from the time of planetary formation.

This traditional concept, however, comes in contradiction with practically all observational evidence. Consider here only a few, most obvious of them (Drobyshevski 1988b, 1997a, 2000, and refs. therein): (*i*) each comet has its own specific signature distinguishing it from others (A'Hearn et al. 2005b); (*ii*) ions and radicals were observed in the coma, in the immediate vicinity of the nucleus, much closer, in fact, than it would follow from the assumption of their photolytic formation from some hypothetical "parent" molecules (particularly remarkable in this respect is the detection of atomic and ionic carbon, C and $C^+$); (*iii*) the origin of CHON dust is unclear; (*iv*) the laws of conservation appear to be violated; indeed, an extremely small fraction of the surface area of the nucleus (≤5-10%) receiving an as small a part of incident solar energy releases large amounts of gases and dust in the form of jets, some of the jets being active on the night side of the nucleus as well; (*v*) origin of the comet outbursts is unclear; moreover, in about 5% of cometary apparitions their nucleus splits into fragments flying away with velocities of up to 1-10 m/s (Sekanina 1982), and so on. This list could be continued further; suffice it to recall the anomalous long-lasting (for weeks!) breakups of P/Shoemaker-Levi 9 into $20^+$ fragments, different in composition, after an encounter with Jupiter with its enormously strong magnetic field, which can hardly be accounted for by tidal effects as this was supposed by some authors (e.g., Asphaug & Benz, 1994; Sekanina et al. 1994) (see discussion in Drobyshevski 1997a), and so on. No physically reasonable and non-contradictory answers have thus far being supplied to these questions. The only thing left to the proponents of the condensation-sublimation concept is a construction of ever new and frequently mutually excluding hypotheses based quite frequently on unobservable factors (amorphous ice, ejection of grains because of the (dirty) ice cracking under thermal impact (with this "impact" lasting years!), conservation in ice of energy-excessive chemical compounds and radicals created by cosmic rays or of bubbles of compressed gas, etc.).

The idea of ejection of comets from planetary bodies dates back to Lagrange and during the 20th century it was actively advocated by Vsekhsvyatski (1967) as an "eruptive concept" appealing to volcanic processes on the Galilean satellites mainly. The New Explosive (Eruptive) Cosmogony of SP comets is based only on one well established electrochemical fact (Decroly et al. 1957; Petrenko & Whitworth 1999), namely, the possibility of electrolytic decomposition of ice in the solid phase (Drobyshevski 1980a; Drobyshevski et al. 1995). Volumetric electrolysis occurred in the massive icy envelopes with embedded carbonaceous and rock inclusions on moon-like bodies of the type of the outer Galilean satellites as they moved originally in the strong ancient magnetic field of planets or of the solar wind, which generated in them currents of up to $\sim 10^2$ MA. The electrolysis products $2H_2 + O_2$ subjected to pressures of up to $p \sim 0.1$-$1$ GPa build up in ice in the form of clathrates, a stable solid solution.

That an oxygen-based clathrate does exist at $T \approx 271$ K and pressures of $\sim 10$ MPa was demonstrated by van Cleef & Diepen in 1965. The oxygen content in a clathrate may amount to one $O_2$ molecule per $\sim 5.67$ $H_2O$ molecules. At $T \approx 120$ K, the oxygen clathrate is stable at $p \sim 20$ kPa (Byk et al. 1980). As for hydrogen, it was believed until quite recently that, similar to He and Ne, it cannot persist in clathrate structures at low pressures and should escape from the ice by diffusion. The situation changed only two years ago, when Mao & Mao (2004) reported existence of a high-pressure $H_2(H_2O)_2$ clathrate that holds 5.3 wt. % hydrogen at $T < 140$ K even at such a low pressure as $p \sim 100$ kPa (the last figures could also shed new light on some other features of cometary activity).

Saturation of ice with the products of electrolysis, $2H_2 + O_2$, up to concentrations of $\sim 15$ wt. % makes it capable of detonation. This is, however, a relatively weak, not a high explosive mixture. The detonation velocity in it is only $D \approx 5$ km/s, with pressure behind the detonation wave front $p_D \approx 5$ GPa (Drobyshevski 1986) (to compare with $D \approx 7$ km/s and $p_D \approx 30$ GPa for a standard TNT-type explosive (Baum et al. 1975)). Therefore, explosion of electrolyzed ice should not bring about crushing of unexploded fragments and loss of the gases dissolved in them. In a greater extent that relates to possibility of solid mineral grain fragmentation.

Detonation can be initiated by a strong enough meteoroid impact. A global off-center explosion of the ice envelope of a moon-like body should shed off a substantial part (10-90%) of the ice (the actual fraction depends on the mass of the body) (Drobyshevski 1980b; Drobyshevski et al. 1994a).

This approach permits one to explain and relate many astrophysical aspects, starting with the origin and properties of asteroids (Drobyshevski 1980a, 1997b; Drobyshevski et al. 1994a) and of many small planetary satellites (Agafonova & Drobyshevski 1985; Drobyshevski 1988a), specific features in the structure and differences of the Galilean satellites (Drobyshevski 1980b), of Titan with its orbital eccentricity and thick atmosphere and Saturn's rings (Drobyshevski 2000, and refs. therein), and ending with comets and the fine features of their manifestations and chemistry (Drobyshevski 1988b). A number of predictions made on the basis of this concept have been confirmed (a thing the traditional hypotheses cannot boast of), while others are still waiting for confirmation (Drobyshevski 2000, and refs. therein). The latest example, - the Cassini-Guygens mission data are indicative indirectly (Tobie et al. 2006) of a presence on Titan of a liquid water mantle under a not very thick ($\sim 1$-$10$ km) ice crust, - the prediction made a quarter of a century ago (see refs. in Drobyshevski 2000). NEC leads to certain conclusions concerning localization of conditions favourable for the origin of life (Drobyshevski 2002), while on the other hand substantiates the priority of exploration of comets for testing the NEC itself and, the last but not the least, argues convincingly for the need of sending missions to Callisto to test the possibility of explosion of its still unexploded ices, which would provide a real threat to the very existence of Mankind (Drobyshevski 1999).

Viewed from the standpoint of NEC, SP comets are solid fragments of surface layers of the exploded icy envelopes. These fragments contain, besides primitive organics and rock inclusions, also $O_2$ and $H_2$, products of electrolysis dissolved in the ice (note that some critics of the NEC (e.g., Shulman 2000), in referring negligently to our publications, retort that $O_2$ and $H_2$ form in the cometary nuclei themselves in their interaction with interplanetary magnetic fields, or that we have in mind volcanic ejections from satellites resulting from $2H_2 + O_2$ explosions in volcanoes; being physically impossible, these processes are not mentioned in our papers at all). A fairly small input of additional energy (through insolation, meteoroid impact etc.) can initiate combustion and even explosions in such ices. One has to bear in mind that, being a product of geochemical differentiation and geological processes on the parent planets, these ices are, as a rule, non-uniform, have inclusions and a layered structure.

An impartial look reveals that all of the available observational evidence, including the above-mentioned facts, which are quite often referred to as "mysterious" and "incomprehensible", finds readily explanation within NEC without invoking any new hypotheses.

Recent data suggest, in particular, that a large part of the fairly well studied nuclei have an elongated shape (for instance, P/Halley, P/Borelli, P/Tempel 1, etc) (Jewitt et al. 2003), a feature characteristic of fragments originating from explosions of much larger bodies with geologically evolved structures (when their fragments of irregular shape are accelerated by a drag of the expanding gaseous products of the explosion more effectively and so escape the parent body easier), rather than of the products of accretion or, conversely, of collisions or collisional erosion. As for the quasi-spherical nucleus of P/Wild 2, the interesting features of its topography, including the irregular shape of the depression with abrupt walls, can be better understood if one takes into account the possibility of local burnout of isolated inclusions and layers enriched in combustible components.

## 3. IMPACT CRATER FORMATION. STRAIGHTFORWARD ESTIMATES FOR TEMPEL 1

The active probing of the cometary nucleus with an impact cratering was aimed at obtaining a basis for a sound judgment of its structure and composition. As a certain measure of our ignorance in this area may serve, for instance, estimates of the possible crater size, $50 < d < 200$ m (e.g., A'Hearn et al. 2005a; Kruchynenko et al. 2005; Hughes 2006), which show that its volume is predicted to within about one to two orders of magnitude. Melosh (1989, Ch. 7) also suggests this figure for the accuracy of crater volume prediction if no experimental data on impacts in the given conditions are available.

Copious literature deals with the formation of craters by impact and explosions (e.g., Stanyukovich 1971; Roddy et al. 1976; Bazilevski et al. 1983; Anderson 1987; Melosh 1989). Straightforward considerations (see, e.g., Bazilevski et al. 1983; Melosh 1989) suggest that the crater volume $V$ should be proportional to the energy $E$ released in an explosion (or impact) and inversely proportional to the energy $q$ absorbed, on the average, by a unit volume of target material:

$V = \chi E/q$,  (1)

where $\chi$ is a coefficient determined empirically ($\chi \sim 10^{-1} < 1$, which is due to one of dissipation mechanisms of many being, as a rule, considered dominant; hence it follows that an efficiency of matter ejection from crater is not very high, - it achieves ~10-30% at best). Based on experimental data, one frequently assumes for the crater depth $h$ vs. diameter $h \approx d/4$, so that $V \approx \pi d^3/32$.

In the energy-based approach, one singles out usually two main modes determining the crater size, namely, the strength and gravitational ones.

We start with the last one. Here the order of magnitude for $q$ is defined as $q \approx \rho g d$ (where $\rho$ is the target material density, and $g$ is the acceleration of gravity), which yields

$$d \approx \left(32 \chi E / \pi g \rho\right)^{1/4}. \tag{2}$$

For $E = 20$ GJ, $\rho = 1000$ kg m$^{-3}$, $\chi = 0.1$, and $g = 10^{-4} g_0$ we obtain $d \approx 400$ m.

If the impact energy is absorbed primarily by plastic deformation of the target material, $q = Y$, i.e., the elastic limit of the material. For rough estimates one usually sets $Y = const$; for granite $Y \approx 100$ MPa, for solid ice $Y \approx 17$ MPa at 257 K, and $Y \approx 34$ MPa for $T = 81$ K (e.g., Lange & Ahrens 1987). Whence, assuming the nucleus of the comet to consist of solid ice at $T = 120$ K (i.e., $Y \approx 30$ MPa), we arrive at

$$d \approx \left(32 \chi E / \pi Y\right)^{1/3} \approx 10 \text{ m}, \tag{3}$$

which is substantially less than the estimate obtained in the gravitational approach.

One can hardly assume, however, that at high strain rate matter would behave as a continuous medium. Only a small fraction of its volume bears the main load, and this reduces strongly $q$. Indeed, in the presence of high shear strain rate gradients adiabatic shear bands appear, which initiates formation of a gas (and even plasma, - see experiments by Drobyshevski et al. (1994b)) phase in the conditions where volume-averaged hydrodynamic consideration would suggest the very onset of a liquid phase formation. This relates to such high-plasticity materials as metals. Therefore, ejection out of craters of solid blocks (by the hot gas component) occurs with a higher velocity and efficiency (Drobyshevski 1995). Another mechanism reducing the effective value of $q$ is the brittle fracture of material by the shock wave that crosses it (Stanyukovich 1971). Here the volume energy expended to crush monolithic material into large blocks is likewise much lower than that needed to shift molecular layers with respect to one another. This is particularly typical of brittle material with a low elastic deformation threshold (ceramics, rocks etc.). As follows from calculations of Nolan et al. (1996), in large scale impacts (impactor size >5 m for $u_{imp} > 5$ km/s) $q$ drops to such low levels that in real conditions crater formation on bodies already as small as ≥1-10 km occurs in the gravitation regime, with ejection of large fragments at low velocities determined by elastic stress relaxation.

There are more sophisticated approaches to estimation of the consequences of impacts (with inclusion of momentum transfer, with the use of the so-called $\pi$ parameters, etc. (e.g. Schultz et al. 2005)). All of them, however, are based on empirical normalizations, which are determined each time for impacts of a given class, and are capable of estimating the crater volume, other conditions being equal, at best to within an order of magnitude (Melosh 1989).

Even the rough estimates presented above demonstrate that the expected size of the crater on Tempel 1 had to be determined by such a large set of unknown parameters and lies, therefore, within such a large range, that even an accurate enough measurement of the crater diameter and depth would hardly permit a reliable judgment of the material and structure of the cometary nucleus or shed light on its composition (for ice - monolithic or porous, amorphous, low or high pressure phases; fraction, state of dispersion, stratification etc. of rock and organic inclusions; presence and structure of nonvolatile crust, angle of impact, and so on). Such conclusions entirely agree with the pre-impact conclusions by Schultz et al. (2005) and Richardson et al. (2005). We might add to this list one more parameter, namely, internal source of energy. The possibility of the presence of an internal energy source at a level of >0.1 MJ kg$^{-1}$ in the comet nucleus was considered by nobody besides us, however it was mentioned by Schultz et al. (2005). Our main hope was placed therefore on an analysis of the composition of gas jet components, their duration, more specifically, on the formation of long-lived gas/dust jets out of the crater or of its impact-perturbed surroundings, measurement of the energetics of the outflows and ejections and, possibly, of large fragments of the nucleus.

## 4 IMPACT CONSEQUENCES WITH INCLUSION OF FORCED DETONATION OF ELECTROLYZED ICES

As already mentioned, all mysterious and bizarre manifestations of comets are readily accountable for by assuming that their ices are saturated (non-uniformly) by the products of electrolysis up to a concentration $\alpha \approx$ 15-20 wt %. Our early estimates (Drobyshevski 1986) suggest that such a uniform solid solution is capable at $\alpha \approx$ 17 wt % of stable detonation at the initial temperature $T_0 \geq$ 145 K, i.e., at a temperature reached in icy envelopes of Galilean satellites at a depth of tens of km. We did not consider reaction kinetics behind the shock front. As a criterion of detonation, i.e., of instantaneous liberation of energy initiated by shock compression of material in the shock wave we accepted the temperature $T_c$ = 900 K reached in the products of detonation (for conventional explosives, the lowest temperature $T_c$ = 700-900 K, see pp. 165-166 and 195-197 in Baum et al. 1975). Therefore, at $T_0 \approx$ 120 K (distance to Jupiter) stable detonation throughout the volume of the body is impossible, as soon as even under the assumption of the reactions being fully completed the temperature would be lower than $T_c$ = 900 K.

At the instant a body hits the ice with $u_{imp} \sim$ 10 km/s the temperature reaches ~10$^4$ K. The energy of DI (~20 GJ) is only enough to melt less than 38 t of conventional ice which was originally at $T_0$ = 120 K. In actual fact, however, this figure will be smaller (see below), because part of the energy in the area of direct contact will be expended to heat material to the plasma state and will be driven by the decaying shock wave into the bulk of the target while only partially transforming into the kinetic energy of directed motion of material. If, however, as *a result of the initial impact* the temperature behind the shock front $T \geq T_c$, the mixture 2H$_2$ + O$_2$ + organics will be able to react with liberation of energy, thus sustaining the shock wave and slowing down its damping. We deal here with the phenomenon of forced detonation (Baum et al. 1975) where liberation of energy behind the shock front is insufficient to make it stationary. In the calculations that follow we have accepted persistence of 2H$_2$ + O$_2$ stoichiometry, bearing in mind the existence at low temperatures and pressures of stable hydrogen and oxygen clathrate hydrates discussed in Sec.2.

Our purpose will in this case be estimation of the additional energy that will be added to 20 GJ of the impact, as well as of the mass that will be involved in forced detonation. The latter figure is important, because it determines the amount of the products of high-temperature exothermal reaction O$_2$ + H$_2$ + organics which hopefully could be estimated in observations of the consequences of the impact. The analysis below was conducted along the lines of the ideology formulated by Drobyshevski (1986).

Interaction of an impactor with the plane surface of the cometary nucleus was numerically simulated in terms of one-dimensional, spherically symmetrical, formulation. We considered evolution of a spherical layer in half-space, with an impactor-initiated pressure pulse specified at a point ($x = 0$) on the surface of this layer.

Propagation of a shock wave was described by standard equations of the mass, momentum, and energy conservation in integral form (Godunov 1976):

$$\oint_\Gamma a dx - b dt = \iint_\Omega \frac{2}{x}(f-b)dxdt, \tag{4}$$

$$a = \begin{bmatrix} \rho \\ \rho u \\ \rho(e+u^2/2) \end{bmatrix}, \quad b = \begin{bmatrix} \rho u \\ p+\rho u^2 \\ \rho(e+u^2/2)u + pu \end{bmatrix}, \quad f = \begin{bmatrix} 0 \\ p \\ 0 \end{bmatrix},$$

where the internal energy of the medium $e$ includes the possible ice-water-vapor phase transitions, as well as the energy released in exothermal reactions behind the shock front.

At this time, one could hardly venture a guess on thermodynamic properties of the solid solution (clathrate?) of $2H_2 + O_2$ in dirty cometary ice, all the more that it contains ~10% of primitive organics. Therefore, to describe the properties of matter behind the shock front and close the system (4), we assumed an equation of state for water in the form (Baum et al. 1975; Drobyshevski 1986):

$$p(\rho,T) = \frac{2.992 \cdot 10^8 (r^{7.3}-1)}{1+0.7(r-1)^4}(1-0.012R^2 f) + 4.611 \cdot 10^5 Rf(T-273), \tag{5}$$

$$f = \frac{1+3.5r-2r^2+7.27r^6}{1+1.09r^6}, \quad r = \frac{\rho}{\rho_*}, \quad \rho_* = 1000 \text{ kg m}^{-3}.$$

The internal energy of water can be written as

$$e_w(\rho,T) = 6.3 \cdot 10^6 \left(1-\frac{1}{r}\right)\left(0.71-\frac{1}{r}\right)r^{4/3}\left[1-2r\exp(-r^2)\right] + 3651.28T + \text{const}. \tag{6}$$

Internal energy of ice was represented in the form

$$e_i = c_v T_i \text{ J}, \tag{7}$$

where the specific heat of ice depends on its temperature as

$$c_v = 7.7 T_i \text{ J kg}^{-1} \text{ K}^{-1}. \tag{8}$$

It was assumed that for pressure $p \geq 2\times 10^8$ Pa the ice-water phase transition temperature can be approximated by the relation

$$T_{iL} = 0.508 \cdot 10^{-7} p + 243 \text{ K}, \tag{9}$$

and the heat of the ice-water phase transition

$$L = 3 \cdot 10^5 \text{ J kg}^{-1}. \tag{10}$$

Oxidation reactions liberate energy at temperatures above a critical level $T_c$, which lies in the 700-900 K interval (Baum et al. 1975) and depends on $\alpha$, the mass content of the products of electrolysis, $2H_2 + O_2$, in the mixture

$$Q = 13.27\alpha \text{ MJ kg}^{-1}. \tag{11}$$

Because the binding energy of the $H_2$ and $O_2$ molecules with $H_2O$ molecules and with one another in ice is small compared to that between water molecules, it was assumed that the $2H_2 + O_2$ components with a mass fraction $\alpha$ are not involved in the energy-consuming phase transitions of water and behave as an ideal gas.

Equations (4) were solved for the region behind the shock wave. It was assumed to propagate with the velocity

$$D = \frac{1}{\rho_0}\left[(p-p_0)/\left(\frac{1}{\rho_0}-\frac{1}{\rho}\right)\right]^{1/2}, \quad (12)$$

where $p_0$ and $\rho_0$ are the initial ice pressure and density (in Drobyshevski 1986, the square brackets were omitted by misprint), and the fluxes at the boundaries of the considered volume were determined by means of the Rankin-Hugoniot relation at the pressure jump, with inclusion of the possibility of a phase transition and an additional energy release:

$$e - e_0 = \frac{p+p_0}{2}\left(\frac{1}{\rho_0}-\frac{1}{\rho}\right). \quad (13)$$

In accordance with NEC, it was assumed, as this was done by Drobyshevski (1986), that the density of cometary ices is that of the high-pressure phases ($\rho_0 = 1183$ kg m$^{-3}$), which persist at low temperatures ($T_0 \sim 100$ K) and after removal of the load, say, at $p_0 = 10^5$ Pa. Decreasing $\rho_0$ to 920-1000 kg m$^{-3}$ affects the result weakly. (The available estimates of the low density of cometary nuclei, $\rho \sim 300\text{-}500$ kg m$^{-3}$, were obtained within pure sublimation models with not taking properly into account the jet-like pattern of outflows from a nucleus; the DI observations of the ejected material permitted to estimate more precisely the real mass and density of the nucleus (see, however, below).)

The amplitude of the initial pressure pulse $p_{\text{imp}}$ can be roughly estimated from the following relation (Baum et al. 1975)

$$p_{\text{imp}} = \frac{\rho_{\text{ice}} c_{\text{ice}} \rho_{\text{cop}} c_{\text{cop}} u_{\text{imp}}}{\rho_{\text{ice}} c_{\text{ice}} + \rho_{\text{cop}} c_{\text{cop}}}, \quad (14)$$

where $\rho_{\text{cop}}$ and $\rho_{\text{ice}}$ are, accordingly, the densities of the impactor (copper) and cometary envelope (ice), $c_{\text{cop}}$ and $c_{\text{ice}}$ are the sonic velocities in these media, and $u_{\text{imp}}$ is the impact velocity. Whence one obtains for the initial pressure in the shock wave $p_{\text{imp}} \approx 40$ GPa.

The pulse duration was estimated as

$$p_{\text{imp}} S_{\text{imp}} t = m_{\text{imp}} u_{\text{imp}}, \quad (15)$$

where $m_{\text{imp}} u_{\text{imp}}$ is the impactor momentum and $S_{\text{imp}} \approx 0.5$ m$^2$ is the impactor area, which yields for the pulse duration

$$t = \frac{m_{\text{imp}} u_{\text{imp}}}{p_{\text{imp}} S_{\text{imp}}} \approx 10^{-4} \text{ s}. \quad (16)$$

Calculation of the processes evolving with time behind the shock wave propagating away from the impact zone into the nucleus of P/Tempel 1 for $\alpha = 0$, as well as for $\alpha = 0.183$, a level high enough to sustain stable detonation ($T_c = 900$ K) at $T_0 = 150$ K, is illustrated in Figs. 1a,b.

We readily see that starting from $x = 1.81$ m from the point of impact for an initial ice temperature $T_0 = 100$ K, and from $x = 1.98$ m for $T_0 = 120$ K, the graphs plotting the variation of the parameters of material ($T$, $\rho/\rho_0$, $D$, $u$) behind the shock front undergo a break, because the shock wave begins to decay rapidly as a result of the excess energy no longer being released. Interestingly, because of a temperature drop due to the energy release ceasing, the density behind the shock wave even increases somewhat initially.

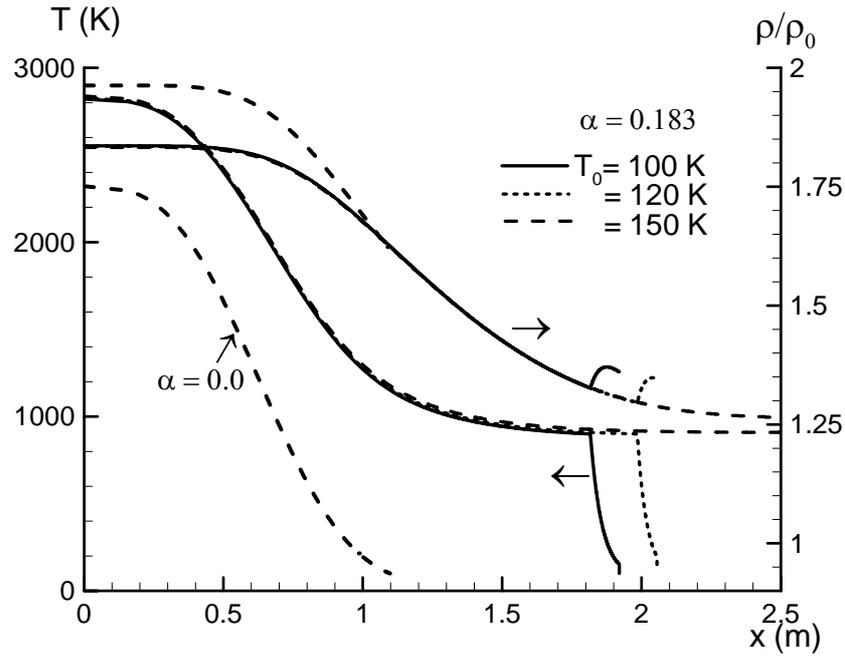

**FIG. 1a.** Temperature $T$ and density $\rho/\rho_0$ behind the shock wave calculated for $\alpha = 0$ and $\alpha = 0.183$ and the original ice temperature $T_0 = 100$, 120, and 150 K and plotted vs. distance $x$ from the point of impact. $T_c = 900$ K.

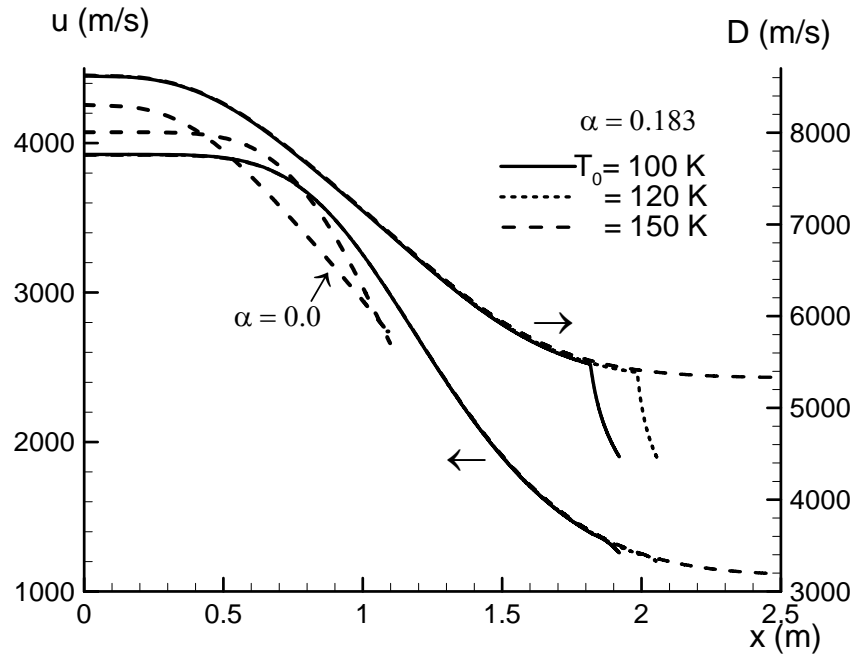

**FIG. 1b.** Shock wave velocity $D$ and velocity u behind the shock wave calculated for $\alpha = 0$ and $\alpha = 0.183$ and the original ice temperature $T_0 = 100$, 120, and 150 K and plotted vs. distance $x$ from the point of impact. $T_c = 900$ K.

Figure 2 provides an idea of the total amount of additional energy liberated in the cometary material as a result of initiation in it of forced detonation. Significantly, this energy, under reasonable assumptions, may exceed noticeably the planned energy of DI.

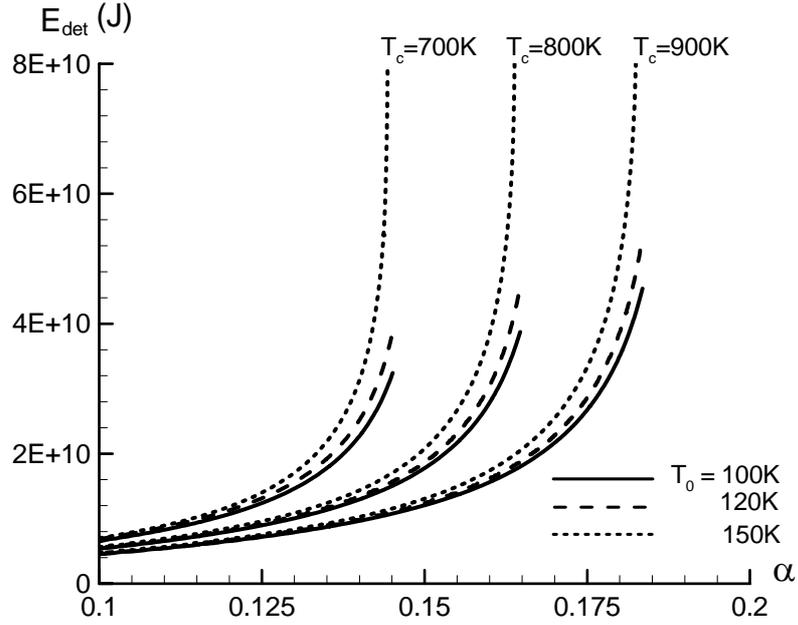

**FIG. 2.** Energy released by the shock wave plotted vs. $\alpha$, the $2H_2 + O_2$ mass content in ice, at various critical detonation temperatures $T_c$ and $T_0 = 100$, 120, and 150 K.

## 5. DISCUSSION OF POSSIBLE OBSERVATIONAL CONSEQUENCES OF THE IMPACT

As follows from the calculations (Fig. 2), the additional energy liberated in the electrolyzed ices of the comet as a result of their impact-initiated forced detonation exceeds the energy of the impact itself only about two- to threefold. This is too small a value to drive the crater size beyond the broad range of uncertainty of the predicted figures.

    The impact itself with $E_0 = 20$ GJ could convert to water <38 t of pure ice (at $T_0 = 120$ K), evaporate <6.2 t, or heat <5.5 t of ice to vapor state at $T = 900$ K. In actual fact, taking into account the strongly non-uniform distribution of energy, part of which would be expended to overheat the material in the immediate vicinity of the impactor and the impactor itself, these figures should be scaled down by about a factor five (indeed, as follows from Fig. 1a, for $\alpha = 0$ only 0.925 t of material would be heated to $T \geq T_c = 900$ K). By contrast, in the case of detonation accompanied by liberation of chemical energy $E = 2.5E_0 = 50$ GJ, when release of additional energy will give rise to a kind of quasi-thermostatting behind the shock wave, the amount of ice, together with the organics it contains, vaporized and inhomogeneously heated to $T \geq T_c = 900$ K will be 19.2 t. This exceeds by more than an order of magnitude (~20 times) the values associated with the impact alone. Therefore, the total (thermal and kinetic) energy contained in the outflow of the gas and of the finely dispersed inclusions and products of some gas components' condensation in expansion in vacuum may turn out comparable to $E_0$.

    Accordingly, the amount of the products of pyrolysis of the organics contained originally in ices should increase by an order of magnitude too.

    What should increase by several orders of magnitude, is the amount of the products of combustion of these organics under deficiency of the oxydizer, which is

determined by the relative amounts of these organics and of the clathrate hydrogen retained in the ices. These are, besides $H_2O$, the various "quasi-cometary" molecules such as $H_2$, $N_2$, $CO$, $CO_2$, $HCN$, $H_2S$, $H_2CO$, $CH_3OH$, light pyrolyzed hydrocarbons etc., as well as, on the one hand, their radicals and ions (however, in relatively low concentrations due to the shock-wave-caused reactions proceed at high pressure), and on the other, carbon-containing soot and CHON particles accounting for the "smoke" produced in incomplete combustion. It is essential that these components should form immediately at the instant of ejection rather than at a large distance from the nucleus, which could be assigned to subsequent photolysis. The existence of a well fixed point of reference, namely, the time of impact, distinguishes favorably this experiment from flyby observations of jets evolving continually from narrow discrete sources on the surface of the nucleus and carrying the low-pressure combustion products containing a great deal of radicals and ions. It would be instructive to monitor the practically instantly ejected cloud of the high-pressure combustion products to see how photolytic and solar-wind-related processes will convert them into a substance of cometary coma and tail which is rich in radicals and ions.

One cannot exclude the possibility that forced detonation will transform to deflagration, i.e., detonation-initiated non-shock low-pressure combustion of material. In this case, the outflow of combustion products from the crater would tail out, and we would become witnesses to (*i*) formation of jets of material emanating even on the night side of the nucleus, and (*ii*) a gradual increase in crater size (not so fast as the one caused by the impact) accompanied by the appearance of burnt-out grooves where the content of combustibles was originally enhanced (recall the Stickney crater on the burnt-out Phobos (Drobyshevski 1988a) and possibly similar large craters on minor low-density bodies (Thomas 1999)).

Our analysis was necessarily restricted to an idealized model of impact on uniform ice. But, first, as already mentioned, cometary ices are products of geological processes in the parent Ganymede-like bodies. Therefore, their structure is spatially non-uniform; indeed, measuring many km in size, they have a complex structure and contain inclusions which were originally enriched or depleted in some minerals, including organics and products of electrolysis. There could even be meter-sized rock inclusions (Drobyshevski 1980a). Second, subsequent evolution of cometary nuclei including loss of volatiles gives rise to development of a non-uniform surface crust consisting of "sand" strengthened by pyrolyzed and cosmic ray processed organics.

It thus appears hardly possible to predict unambiguously the consequences of the impact. Impact on a boulder would bring about results radically different from those of an impact on a vein with an enhanced concentration of the products of electrolysis and organics, somewhere in the vicinity of a jet source. In the latter case, which would be of most interest for a scientist, one could conceive of a situation where waveguide properties of such a vein would drive detonation very deep into the nucleus and even culminate in its breakup into large fragments flying away from one another with $E \sim E_0$ or even greater, a case observed on many occasions.

## 6   COMPARISON OF DI EXPERIMENT WITH CONCLUSIONS OF THE OLD PARADIGM AND THE NEC CONCLUSIONS

Now turn to interpretation of the DI experiment results following the conclusions and recommendations made in the previous Section. First, it is necessary to note that the treatment of the obtained DI data was conducted by the authors of the experiment from positions of the traditional paradigm. Therefore, when the facts correspond to an expected scheme in not entire extent, the interpretation is often accompanied by words of a kind: "it is probably associated with…", "this is undoubtfully due to…", etc.

First of all, it is possible to note with satisfaction that the mean nucleus density is evaluated by the DI team as $\rho = 620$ (+470/-330) kg/m$^3$ (A'Hearn et al. 2005b). That, generally speaking, exceeds the initially accepted magnitude 500 kg/m$^3$ for the working core model (Belton et al., 2005), and conflicts with the concept on the nucleus as a lump of dirty loose snow (which, it seems as if, is confirmed by an unexpectedly large quantity of the matter ejected from the crater) but, however, corresponds to the model of a monolithic nucleus that follows from NEC (in Sec. 4 we took $\rho \approx 1180$ kg m$^{-3}$ for the calculations). Unfortunately, it turned out to be impossible to measure the crater size (and, probably, in future it remains impossible too (Hughes 2006)) due to the impact site passed behind the horizon relative to flyby module to the time of scattering of the matter ejected. Nevertheless, the first impression of the DI team was such that the ejected mass turned out unexpectedly large. The shadow thrown by the plume evidences that its base diameter was >300 m, which "much wider than the expected size of the crater at this early stage" (A'Hearn et al. 2005b). If the crater size really exceeds the expected one, then at $\rho \rightarrow 1000$ kg m$^{-3}$ this may evidence a monolithic (under not very thick porous surface layer) nucleus containing products of electrolyze and organic matter. If an excess detonation energy release takes place, then, obviously, the crater size can not be determined unambiguously only by the impact energy and strength of the target matter. The second strong burst shifted downward, that was observed in ~200 ms and resulted in saturating the detectors, can be interpreted as the initiation by the impact of detonation in the nearest nest with a heightened concentration of $H_2+O_2$ and organics.

The density $\rho \approx 620$ kg m$^{-3}$ mentioned above should be considered as the lower estimate. It was evaluated under an assumption that the observable duration of the existence of (nontransparent) ejection (~0.5 hour) is conditioned, basically, by the time of ballistic flight of particles ejected during the first seconds after the collision. If in reality a longer efflux of the matter from the crater takes place, as it follows from NEC (see Sec. 5), and is seen in DI images (Fig. 9 in paper by A'Hearn et al. (2005b) shows that this time > 45 minutes), then the estimate of the nucleus mass (and its density) increases in obvious way and nears to $\rho \approx 1000$ kg m$^{-3}$, and may be even somewhat higher.

A'Hearn et al. (2005b) modeled a nontransparent self-luminous plume as "4000 kg cloud of liquid silicate droplets of 150 μm diameter…, expanding at 1.7 km s$^{-1}$, and with initial temperature 3500 K". Such a mass of silicates with their mean specific heat 1230 J kg$^{-1}$K$^{-1}$ should contain about 17 GJ of heat plus 6 GJ of kinetic energy. That exceeds the impact energy and does not leave an energy fraction comparable in magnitude for other plume components (slower silicate and carbon dust with a mass of ~ 10$^6$ kg, see below) and for shock waves having propagated inside the target. We neglect that the silicates decompose at $T > 3000$ K and, according to our calculations (see Sec. 5), the impact is capable of heating less than 1000 kg of water initially being ice up to $T \geq 900$ K with no additional energy release. It would be more natural to assume that the optical plume properties are governed not by silicate particles but submicron particles of amorphous carbon that was discovered in products of ejection from DI crater by Harker et al. (2005) with the help of the IR spectroscopy. The amorphous carbon exists up to 4000 K and just it, according to Meech et al. (2005), governs the dust temperature. This is soot that, similar to CO, is a product of incomplete combustion of an organic matter at deficiency of the oxidant which is mentioned in Sec. 5, and as we repeatedly noted in the past when discussing manifestation of the comet activity (Drobyshevski 1988b). The same concerns also other light molecules containing carbon like $CH_4$, $C_2H_6$, $CH_3OH$, $C_2H_2$, HCN. They appear due to pyrolysis of heavier organic molecules.

Some words on the plume temperature. In our present calculations, as in 1986, we assumed that the detonation occurs when the temperature attains $T_c \approx 900$ K. This magnitude well corresponds to the observable dust temperature 850 K in the plume, to the lower limit (1000 K) of excitation of rotational freedom degree of $H_2O$ and $CO_2$ in the plume (A'Hearn et al. 2005b), as well as to the observed velocity 5 km s$^{-1}$ of expansion into vacuum of the leading front of the plume if molecules of $H_2O$ are main gas component of the plume (the velocity of the front boundary of a gas expanding into vacuum is defined by expression $u_{max} = [2/(\gamma-1)](\gamma kT/\mu)^{-1/2}$ where $\gamma = 1.32$ is the adiabatic exponent for $H_2O$, $\mu$ – is the molecular weight, $T$ – is the initial gas temperature (Zel'dovich & Raizer 1966)). After the impact, in the dust, in parallel with amorphous silicates that appeared when condensing from the pre-planetary gas phase and were the main dust component in the pre-impact cometary coma, a crystalline phase was also detected by Lisse et al. (2005) and by Harker et al. (2005). This crystalline phase amounts to a third of the entire silicate dust mass.

The amorphous phase of silicates transforms into crystalline one when being heated up to ~1000 K but again. According to Sugita et al. (2005) estimates, the total impact energy is sufficient for crystallization only of 20 t of silicates, while the dust contains no less than 180 t of crystalline pyroxene and olivine. On this basis, the authors suggest a new hypothesis *ad hoc* on that "a substantial amount of the material… went through high-temperature conditions in the early solar nebula" (Harker et al. 2005; Sugita et al. 2005), although, repeat, before the impact the dust contained in detectable amount only the amorphous olivine (Harker et al. 2005). Just NEC proposes the temperature behind the front of the detonation wave that is needed for the crystallization.

The total quantity of the dust ejected as a result of the impact is evaluated as ~$10^6$ kg at its expected velocity ~200 m s$^{-1}$ (Meech et al. 2005) or (5.6-8.5)x$10^5$ kg at 130 m s$^{-1}$ (Harker et al. 2005; Sugita et al. 2005). Therefore its kinetic energy is practically equal to the energy of DI impact, whereas, in the best case, it may amounts to about 10-30% if one takes into account other energy withdrawals.

The detection of spectral evidence of the presence of carbonaceous materials (carbonates and hydrogenated aromatic hydrocarbons) in the dust (Lisse et al. 2005), as well as the heterogeneous and layered structure of the nucleus itself, also evidence in favour of geological evolution of the nucleus matter in the ice (water) mantle of the parental planet body (Secs. 2 & 5; Drobyshevski 1997a).

We avoid to speak here of a large excess of $H_2O$ molecules in the coma of comet Tempel 1 after the collision (more than 150 t which is much larger than the DI energy is capable to vaporize; this quantity follows from the number of $H_2O$ molecules added to the coma after the impact (~$5\times10^{30}$, see Table 1 in Mumma et al. 2005) in the limits of pencil beam of ~140 km in radius (spectrometer setting KL1); the molecules in these limits renew each ~ 5 min, so the number mentioned should be increased probably by 4-5 times, that is, up to ~ 600-700 t). We also let alone a faster increase of the coma brightness in 7 min after the collision (Meech et al. 2005). In principle, one may try to explain the both cases by evolution (sublimation and reduction of the optical thickness in the course of expansion) of the cloud of the ejected dust and ice grains.

Note, on the other hand, that the DI-caused event, probably, slightly differ from natural outbursts observed before and after DI (A'Hearn et al. 2005b; Meech et al. 2005). The projected expansion velocity of the produced dust clouds has been ~200 m/s. The time of growth of the brightness amounts to a few minutes (from this stand point, observations of "distinct rates of brightening… in the first few minutes" of the DI-stimulated event (Meech et al. 2005) are of interest), and the sources of some of them are frequently associated with the same place on the nucleus surface (A'Hearn et al. 2005b) (here it is reasonable to put a provocative question: "What, besides NEC with its natural and physically transparent ideology, is capable of explaining these phenomena?"). Comparison of the consequences and energy release (tens of GJ) of the DI event with the natural outbursts offers also an answer to the question "why a major jet did not occur after the excavation of a volatile-rich layer?" (Sugita et al. 2005), although, if one remains in the frames of the former concepts, volatiles probably sublime below the surface (A'Hearn et al. 2005b). We give an answer at the end of Sec. 5. Unfortunately, the most interesting possibility purely accidentally was not realized in this experiment: DI impactor did not strike the extended layer (scarp or vein) excessively enriched by the electrolysis products from that the natural outbursts emanate, and so the nucleus did not split. Nevertheless the impactor got a place where, conventionally speaking, $\alpha \rightarrow 0.183$ (this number depends on $T_0$, see Sec. 4). Just for this reason, DI experiment showed though (*i*) a number of indications of an excessive energy release (and heating of the matter) in a regime of forced detonation considered in Sec. 4, and as a consequence, (*ii*) evidence of an outflow from the newly formed crater lasting minimum during ~1 hour, which hardly may be explained by the solar radiation falling into the crater, because the radiation has to be screened by the non-transparent efflux products.

## 7. CONCLUDING REMARKS

When reading papers discussing the results of DI experiment, it is difficult to overcome an impression that their authors following traditional approaches concerning the comet nature undergo noticeable difficulties when they try to coordinate observations with the conservation laws. It is instructive that these challenges vanish if one is based on NEC.

A simple example. Twenty and ten seconds before the impact (that is, ~ 200 and 100 km distant from the nucleus) the DI impactor suffered two collisions with large dust particles that disturbed its orientation (A'Hearn et al. 2005b). (Remind that analogous events, that destabilized Giotto probe of 574 kg mass in 1986, were caused by hypervelocity ($u_{imp} \approx 68$ km s$^{-1}$) impacts by particles with the masses ~40 and 8 mg at distances ~3000 – 1000 km from the P/Halley nucleus (McDonnell et al. 1986).)

During 10 s the DI impactor sweeps a volume of $10^4$ m$^3$, which corresponds to the concentration ~$10^{-4}$ m$^{-3}$ of the particles mentioned, so that at the dust velocity ~100 m s$^{-1}$ relative to the nucleus this corresponds to the dust flux $10^{-2}$ m$^{-2}$s$^{-1}$. The DI event caused a liberation of the dust mass equivalent to ~10 hours of normal pre-impact dust production (Meech et al. 2005) (this was a dusty impact). The most part of the dust mass of the DI ejection is concentrated in the largest particles. During 10 hours under a spherically symmetric efflux the comet loses ~$4.5 \times 10^{13}$ such particles, which results in $\geq 4.5 \times 10^7$ kg of the total dust mass. Harker et al. (2005) and Sugita et al. (2005) in their estimation of the ejected DI mass when integrating the Hanner grain size distribution with respect to particle sizes confined themselves by a somewhat arbitrarily accepted upper bounds of the particle sizes from 1 to 100 μm (or, probably, they tried to meet the conservation laws?). In this case, it is possible (although barely) to agree the kinetic energy of the dust with the DI energy. However, our moderate estimate of the dust mass obtained just now exceeds their estimates as a minimum by 1.5 order. The contradiction may be hardly overcome without an additional energy release in the crater. Naturally, it is necessary to refine the mass of the particles, destabilizing the impactor, however, probably, a particle with the mass <1 mg will be incapable of doing much. The probability is that masses of these two particles, with regard for their relatively low velocities (10 km s$^{-1}$ as compared 68 km s$^{-1}$ for Giotto) and, consequently, smaller momentum transferred to the DI impactor, exceed 1 mg noticeably. (Note that the momentum gained by the target at a hypervelocity impact considerably exceeds the product of the impactor mass by its velocity due to the large momentum of the recoil caused by the high-velocity ejections from the crater (Drobyshevski et al. 1994b); the recoil contribution increases faster than the impactor momentum.) As like as not that a correction should be made taking into account non-sphericity of outflows and ejections from the pre-impact nucleus etc, which, however, hardly influences critically on the particle distribution in size and their maximum size. Nevertheless, it is possible to state that the condensation-sublimation comet paradigm widely distributed now demands additional hypotheses for treatment many facts, and for this reason this paradigm now really serves an obstacle on the way of comprehension of the true nature of comets and their activity. (What is a worth, for example, discourses on difference in the nature of SP and LP comets having been investigated in detail in recent years (e.g., Harker et al. 2005; Sugita et al. 2005; Mumma et al. 2005).)

As for LP comets, their origin somewhat differs from one of SP comets. From close-binary cosmogony of the Solar system it follows that their ice nuclei arise when colliding between distant ice Pluto- and Moon-like planets – members of a common planet-cometary cloud (Drobyshevski 1978). According to our concept, this cloud contains up to $10^3$ planets moving, as a rule, in strongly tilted eccentric orbits, so that this cloud really is a toroid located in the limits ~50-300 au with the strongly diffuse external boundaries. These planets had run to there from the Jupiter zone due to mutual perturbations of initially ~$10^4$ similar objects and due to perturbations exerted on them by giant planets, as well (now add after our not completed attempts to simulate the origin of this planet-cometary cloud) by much larger in mass non-uniformly distributed gas – the product of early evolution of the Jupiter-Sun system when proto-Jupiter yet (over)filled its Roche lobe (at the polytrope index ≈ 3/2).

The improvement of the observation technique during the last quarter of century gave rise to serial discoveries of the planet - members of this toroid-like cloud (e.g., 2003 VB$_{12}$ Sedna, 2003 UB$_{313}$ Xena and others) which confirms our concept and predictions based on it by the facts.

It is obvious that such a planet-comet toroid is neither a <u>narrow comet belt</u> (*sic!*) of Kuiper-Edgeworth at a distance of 40-50 au (these scientists spoke nothing about the presence of many little planets at such distances), nor quasi-spherical Oort comet cloud. If hundreds of icy planets are really contained in this toroid, then their very rare collisions inevitably lead to ejection of ice fragments – nuclei of LP comets. LP comets must differ in their behavior from SP comets at least by that their ices do not contain a large quantity of the electrolysis products and due to this, for example, their maximum activity takes place before the perihelion passage. Of course, one has to remember that mixing of SP and LP objects occurs due to variation of their orbits perturbed by planets, as well as because of transfer (practically also due to planet perturbations) of a part of

comets – fragments resulted from explosions of inner ice bodies, say, Titan, at once to LP orbits soon after their emergence (Drobyshevski 2000).

Thus, from the not anywhere completed analysis conducted in the present paper of the available data of DI experiment it is seen that considerable part of comets may be hardly considered as a debris remained after the planet formation. They are fragments ejected from planets suffered long geological evolution which explains with no contradictions their various properties and behavior, including ones observed in DI experiment.  It is possible only to be astonished, how naturally all the facts seeming to be non-correlated (the masses of particles destabilized the impactor, and, as a consequence, unexpectedly large total mass of the ejection, the minimal detonation 900 K temperature and the temperature of the primary plume, rotational temperature of molecules in it, 5 km s$^{-1}$ velocity of the efflux of the leading plume front, temperature of crystallization of silicates, and, on the contrary, emergence of amorphous carbon - soot and CO, $CH_4$, and many other observations) compose a unified picture of consequences of DI and their connection with many unclear till now manifestations of the comet activity (for example, outbursts and splittings of nuclei, appearing C, $C^+$ near them, etc.), if one proceeds from a simple and physically transparent assumption on inevitability of the large-scale electrolysis in icy envelopes of distant moon-like bodies and, again, from its obvious far-reaching implications.

**ACKNOWLEDGMENTS**

The authors are grateful to M.F.A'Hearn, M.J.S.Belton, D.E.Harker, K.J.Meech, M.J.Mumma, and S.Sugita for providing with reprints on the Deep Impact Project, to K.I.Churyumov for fruitful discussions, and to R.Horgan for drawing our attention to works by W.L.Mao and H.Mao on hydrogen clathrates.